\documentstyle[12pt,epsfig]{article}

\newlength{\dinwidth}                                                         
\newlength{\dinmargin}                                                         
\def\lapproxeq{\lower .7ex\hbox{$\;\stackrel{\textstyle <}{\sim}\;$}} 
                                                         
\def\gapproxeq{\lower .7ex\hbox{$\;\stackrel{\textstyle >}{\sim}\;$}} 
 
\def\be{\begin{equation}}    
                                                      
\def\ee{\end{equation}}                                                         
\def\bea{\begin{eqnarray}}     
                                                     
\def\eea{\end{eqnarray}}

\begin{document}                                                         
\begin{flushright}                                                         
DTP/00/38 \\                                                         
RAL-TR-2000-028\\
July 2000 \\                                                         
\end{flushright}                                                         
                                                         
\vspace*{2cm}                                                         
                                                         
\begin{center}                                                         
{\Large \bf Estimating the effect of NNLO contributions} \\                
                
\vspace*{0.5cm}                
{\Large \bf on global parton analyses}

\vspace*{1cm}                                                         
A.D. Martin$^a$, R.G. Roberts$^b$, W.J. Stirling$^{a,c}$ and 
R.S. Thorne$^d$  \\                                                         
                                                        
\vspace*{0.5cm}                   
$^a$ Department of Physics, University of Durham, Durham. DH1 3LE \\ 
                                                        
$^b$ Rutherford Appleton Laboratory, Chilton, Didcot, Oxon.  OX11 0QX \\ 
            
$^c$ Department of Mathematical Sciences, University of Durham, 
Durham.  DH1 3LE \\ 
$^d$ Jesus College, University of Oxford, Oxford.  OX1 3DW \\ 
\end{center}                                                         
                                                         
\vspace*{1cm}                                                         
                                                         
\begin{abstract}                                                         
We use the recent estimates of NNLO splitting functions,  
made by van Neerven and Vogt, to  
perform exploratory fits to deep inelastic and related  
hard scattering data.  We investigate the  
hierarchy of parton distributions obtained at LO, NLO and NNLO, and,  
more important, the  
stability of the resulting predictions for physical observables.   
We use the longitudinal  
structure function $F_L$ and the cross sections $\sigma_W, \sigma_Z$  
for $W$ and $Z$  
hadroproduction as examples.   
For $F_L$ we find relatively poor convergence, with  
increasing order, at small $x$; whereas $\sigma_{W,Z}$  
are much more reliably predicted. 
\end{abstract}                                               
   
\section{Introduction}   
With the increased precision of deep inelastic scattering data 
\cite{HERA},  
and the need for  
accurate predictions at the Tevatron and the LHC, it is clearly  
essential to extend global  
parton analyses to next-to-next-leading-order (NNLO) in $\alpha_S$.   
Although the relevant  
deep inelastic coefficient functions have been known for some  
time \cite{CF}, there is only  
partial information on the corresponding splitting functions.   
The $N = 2, 4, 6, 8$ (and 10 for  
non-singlet) moments have been calculated \cite{MOM},  
which effectively provide  
information on the high $x$ behaviour of the splitting functions.   
Also known is the most  
singular $\log 1/x$ behaviour at small $x$, both for the singlet \cite{SX},
and the phenomenologically less important nonsinglet \cite{NSSX} splitting
functions, 
and the leading $n_f$ contributions  
\cite{GRACEY1} of the nonsinglet splitting functions, and of the $C_A$
dependent part of $P_{gg}$ \cite{GRACEY2}.   
Recently van Neerven and Vogt \cite{NV} have  
constructed compact analytic expressions for the splitting functions  
which represent the  
fastest and the slowest evolution that is consistent with the above  
information.  We believe  
that these two extreme behaviours are indeed realistic.   
Although there are indications that the  
true behaviour of the splitting functions is likely to be slightly  
nearer to that corresponding to  
the slow evolution possibility\footnote{A view confirmed by private  
communication with A. Vogt.}, for simplicity we shall use the average 
of the two extremes for our \lq central\rq\ NNLO analysis. 
 
It is important to stress an important difference between our  
analysis and the procedure used  
by van Neerven and Vogt \cite{NV}.  The latter authors start from  
a {\it fixed} set of partons and a fixed scale ($\sim 30$~GeV$^2$ i.e 
$\alpha_S = 0.2$) 
and present the differences between LO, NLO and NNLO evolution.   
Here we compare the  
partons, and the consequent predictions for physical observables,  
obtained by performing  
global analyses at LO, NLO and NNLO.   
Both works present NNLO results obtained using  
the extreme estimates of the ${\cal O} (\alpha_S^3)$ splitting functions. 
 
In Section~2 we discuss the changes to the global analysis that are  
necessary in going from a  
NLO to NNLO formulation.  Then, in Section~3, we present seven new 
fits to the deep  
inelastic and related data; that is LO, NLO and five NNLO analyses.   
To gain insight into the  
impact of the NNLO contributions, we discuss essential features of the 
fits in terms of the  
behaviour of the splitting (and coefficient) functions.   
In Section~4 we compare the partons  
obtained in the LO, NLO and NNLO analyses, paying particular  
attention to the gluon  
distribution in the small $x$ region.  The parton distributions are  
scheme dependent and are  
not themselves observable.  The comparison of LO, NLO and NNLO  
predictions for physical observables is much more meaningful.   
In Section~5 we study the predictions for the  
longitudinal structure function, $F_L$.  This is a particularly  
relevant observable as it directly  
reflects the behaviour of the gluon distribution at small $x$, and  
hence most directly probes the stability, or convergence, of parton  
analyses as we go from the LO, to the NLO, and then to the  
NNLO framework.  In Section~6 we compare the LO, NLO, NNLO predictions 
for the cross sections of $W$ and $Z$ boson production at the  
Tevatron $p\bar{p}$ collider and at the  
LHC.  These observables mainly depend on the quark distributions in  
the region $Q^2 \sim 10^4$~GeV$^2$, and $x \sim 0.05$ and 0.006 
respectively. The stability of the predictions  
offers the possibility of using the $W$ and $Z$ events as a luminosity 
monitor of the collider. Finally in Section~7 we give our conclusions. 
 
\section{Global analyses at NNLO} 
 
The procedure is based on the NLO analyses described in  
Refs.~\cite{MRST1,MRST2}.   
However at NNLO it is important to allow the gluon distribution to  
become negative in the low $x$, low $Q^2$ domain.  We therefore adopt 
the parameterization 
\be 
\label{eq:a1} 
xg (x, Q_0^2) \; = \; A_g \: x^{-\lambda_g} \: (1 - x)^{\eta_g} \:  
(1 + \varepsilon_g \sqrt{x} +  
\gamma_g x) \: - \: A_g^\prime \: x^{- \lambda_g^\prime} \:  
(1 - x)^{\eta_g^\prime} 
\ee 
at the starting scale $Q^2 = Q_0^2 = 1~{\rm GeV}^2$ of the evolution.
The parameter $\eta_g^\prime$ turns out to be large in the additional 
negative term and so this contribution is only important at small $x$. 
 
It is necessary to implement other extensions of the formalism when 
going to NNLO.  First,  
we use the three-loop expression for $\alpha_S$,  
in the $\overline{\rm MS}$ scheme.   
Second, we require more detailed matching conditions when evolving  
through the heavy flavour thresholds.  The NNLO treatment of heavy  
flavours is discussed in the Appendix. 
 
Our main interest is in the quality of the fit to deep inelastic data  
at small $x$.  At high $x$  
we have a slight inconsistency in our NNLO analyses in that we use  
NLO expressions to fit to  
Drell-Yan, jet production and $W^\pm$ boson rapidity asymmetry.   
The NNLO corrections to all these quantities have not yet been calculated. 
However note that the  
physical observables that we study (namely $F_L$ and $\sigma_{W,Z}$)  
sample low $x$ partons, which are determined mainly by deep inelastic 
data for which the NNLO formalism is consistent. 
 
\section{The new global fits} 
 
We perform LO, NLO and NNLO global fits to the set of deep inelastic  
and related data that  
was used in Refs.~\cite{MRST1,MRST2}, except that now we use the  
jet $E_T$ distribution measured at the Tevatron to pin down the gluon  
distribution at large $x$, instead of prompt  
photon hadroproduction.  The QCD description of the latter process has 
outstanding theoretical problems \cite{GAMMA}.  A second change is 
that we include all the available  
preliminary HERA data \cite{HERA}, which have higher precision than hitherto. 
 
The consequence of replacing prompt photon data by the jet data is 
that the NLO fit is now similar to that achieved by the previous  
MRST($g\!\!\uparrow$) set of partons \cite{MRST1,MRST2}.  A satisfactory  
description of the Tevatron jet data is obtained, including 
particularly the normalization. 
 
Five NNLO fits were performed.  The \lq central\rq\ fit and the four  
extremes ($A_q A_g, A_q B_g, B_q A_g, \linebreak B_q B_g$), where  
$A_i (B_i)$ corresponds to the slow (fast)  
evolution of parton $i = q, g$.  It turns out that the NNLO fits with 
slow and fast gluon evolution are very similar, and so it is 
sufficient to present results for just two of the extreme  
choices of the splitting functions, namely 
\bea 
\label{eq:a2} 
A & \equiv & A_q A_g~~~({\rm slow~evolution}), \nonumber \\ 
B & \equiv & B_q B_g~~~({\rm fast~evolution}).  
\eea 
 
In Figs.~1 and 2 we show the LO, NLO and NNLO descriptions of the $F_2$ 
data \cite{F2} in a few representative $x$ bins.  
We display only the \lq central\rq\ 
NNLO fit.  However the quality  
of all the NNLO fits is similar.  It is encouraging to note that,  
as we proceed from the LO $\rightarrow$ NLO $\rightarrow$ NNLO 
analysis, there is sequential improvement in the  
overall quality of the description of the data.  In particular, in 
going from the NLO  
$\rightarrow$ NNLO fit, there is an improvement in the simultaneous  
description of the NMC and HERA $F_2$ data.  Indeed the quality of the 
NNLO fit is improved for almost all subsets of the data. 
 
{}From Figs.~1 and 2 we can see that at NNLO the scaling  
violations increase both at small $x$ and at  
large $x$.  At small $x$ this is due mainly to the NNLO contribution 
to $P_{qg}$, whereas at large $x$ the NNLO term in the coefficient  
function plays the dominant role.  The  
relevant $x \rightarrow 0$ behaviour of the splitting functions  
are\footnote{For the LO splitting function  $P^{(0)}_{qg}(x)$  
we use the coefficient of the moment space expression in the limit  
$N \to 0$ rather than the real limit as $x \to 0$.}  
\bea 
\label{eq:a3} 
P_{qg} (x) & = & 2n_f\frac{\alpha_S}{2\pi} \frac{1}{3}\biggl(1 
+\frac{5}{3} 
\frac{\bar \alpha_S}{x} + \frac{14}{9} \frac{\bar \alpha_S^2 \ln(1/x)}{x}  
+ \cdots \biggr)\\ 
\label{eq:a4} 
P_{gg} (x) & = & \frac{\bar \alpha_S}{x}\left [1-\frac{61n_f}{36} 
\frac{\bar\alpha_S}{x}- \biggl(\frac{395}{104} -\frac{1}{2}\zeta(3)  
-\frac{11\pi^2}{72} +n_f\biggl( \frac{295}{2808}-\frac{\pi^2}{972}\biggr) 
\biggr)\frac{\bar \alpha_S^2\ln(1/x)}{x}\right ], 
\eea 
where $\bar \alpha_S =(3/\pi)\alpha_S$, and the $x \rightarrow 1$ 
behaviour of the quark contribution to the $F_2$ coefficient function is 
\bea 
\label{eq:a5} 
& & C_{2,q}(x) \; = \; \delta(1-x) +\frac{\alpha_S}{4\pi}C_F\biggl( 
\frac{4\ln(1-x)-3}{1-x}\biggr)_{+} \nonumber \\ 
& & \quad + \; \left (\frac{\alpha_S}{4\pi}\right )^2 \left [8C_F^2 
\left (\frac{\ln^3(1-x)}{1-x}\right )_{+} 
+ \left (-\frac{22}{3}C_A C_F+\frac{4}{3} C_F n_f -18 
C_F^2 \right )\left (\frac{\ln^2(1-x)}{1-x}\right )_{+}\right ]. 
\eea 

As well as the improvement in the quality of the fit, we can
investigate the importance of the increased scaling violations by looking at 
the higher-twist component of $F_2$   
extracted using a phenomenological analysis in which a term  
$(D (x)/Q^2)F_2(x,Q^2)$ is included in the fit, as in Ref.~\cite{MRST3}.
The values of the higher-twist coefficient $D(x)$ can be seen in Table~1.
At very high $x$ a large positive higher-twist contribution is clearly needed.
This decreases slightly as we move from LO to NLO to NNLO, but there is no 
indication that its presence will be eliminated by even higher orders. 
We note that the conclusion that NNLO contributions largely remove the need for
higher twist at high $x$ in previous NNLO analyses \cite{KATAEV} has been 
based on analysis of CCFR data only, which exists at far higher $W^2$ than 
the SLAC data included in our higher-twist fit, though it has also been
suggested that when NNLO coefficient functions are used the higher twist may be
almost entirely due to target mass effects \cite{BODEK}. At $x =0.4 \to 0.5$
the higher-twist contribution changes sign, becoming generally negative. 
At LO its magnitude is then quite large, 
demonstrating that the evolution is 
too slow at low $Q^2$, both for NMC and HERA data, as is obvious from Fig. 1. 
The magnitude of the higher-twist contribution for $x<0.3$ decreases 
significantly going to 
NLO, and decreases again, to very small values, 
at NNLO. Indeed, the sign of the small-$x$ higher-twist contributions at 
NNLO is not 
even well-determined, with many $x$-bins preferring a slightly positive value.  
The implication seems to be that higher-twist contributions at small 
$x$ are small, and their apparent size is decreased by the inclusion of more  
perturbative corrections.   
    
\begin{table}[h]
\caption{Values of the higher-twist coefficient $D(x)$ extracted from the LO,
NLO and NNLO fits}
\begin{center}
\begin{tabular}{|c|c|c|c|} \hline
& & & \\
$x$ & LO & NLO & NNLO \\ \hline 
0 -- 0.0005 &    $-$0.4754 &   0.0116 &   $-$0.0061 \\
0.0005 -- 0.005 &    $-$0.2512 &   $-$0.0475 &   0.0437 \\
0.005 -- 0.01 &     $-$0.2481 &   $-$0.1376 &    $-$0.0048 \\
0.01 -- 0.06 &     $-$0.2306 &    $-$0.1271 &   $-$0.0359 \\
0.06 -- 0.1 &    $-$0.1373 &   $-$0.0321 &     0.0167 \\
0.1 -- 0.2 &    $-$0.1263 &    $-$0.0361 &    0.0075 \\
0.2 -- 0.3 &    $-$0.1210 &   $-$0.0893 &    $-$0.0201 \\
0.3 -- 0.4 &    $-$0.0909 &   $-$0.1710 &    $-$0.1170 \\
0.4 -- 0.5 &      0.1788 &    $-$0.0804 &    $-$0.0782 \\
0.5 -- 0.6 &      0.8329 &    0.3056 &    0.1936 \\
0.6 -- 0.7 &      2.544  &    1.621  &    1.263  \\
0.7 -- 0.8 &      6.914  &    5.468  &    4.557 \\
0.8 -- 0.9 &     19.92   &   18.03   &   15.38  \\ \hline
\end{tabular}
\end{center}
\end{table}
 
For each fit -- LO, NLO, NNLO -- we use the one-, two-, three-loop expression
for the $\beta$ function, e.g. in the NNLO fits the connection between 
$\alpha_S$ and $\Lambda_{\overline{\rm MS}}$ involves $\beta_2$ evaluated in 
the $\overline{\rm MS}$ scheme. For completeness 
we show in Table~2 the values of the QCD coupling, together with
$\Lambda_{\rm QCD}$, found in the different global fits. For the NNLO fits the
value of $\Lambda_{\rm QCD}$ is kept the same for the extremes as for 
the central
fit, but would change by only a tiny amount if left free. In the fits where 
a higher-twist component is allowed, at each order the extracted value of 
$\Lambda_{\rm QCD}$ increases, reflecting the effect of the increased scaling
violation by the new data at low $Q^2$ included in these fits. This 
increase in $\Lambda_{\rm QCD}$ is only $10-15\%$ (decreasing with
increasing order), leading to a corresponding increase in $\alpha_S(M_Z^2)$ of
$0.002-0.003$.      
 
\begin{table}[h] 
\caption{The QCD coupling and $\Lambda$ parameter} 
\begin{center} 
\begin{tabular}{|l|c|c|} \hline  
& & \\ 
& $\alpha_S (M_Z^2)$ &  
$\Lambda_{\rm{LO\;\; or\;\;} \overline{\rm MS}}^{(n_f = 4)}~({\rm GeV})$ \\  
& & \\ \hline 
LO & 0.1253 & 0.174 \\ 
NLO &  0.1175 & 0.300 \\ 
(NNLO)$_{\rm central}$ & 0.1161 & 0.242 \\ 
(NNLO)$_{\rm A}$ & 0.1161 & 0.242 \\ 
(NNLO)$_{\rm B}$ & 0.1161 & 0.242 \\ \hline 
\end{tabular} 
\end{center} 
\end{table} 
 
\section{Implications for parton distributions} 
 
In Fig.~3 we compare the parton distributions found in the  
NNLO fit to those in the NLO  
analysis.  We plot the NNLO/NLO ratios for the gluon,  
and the up and down quark distributions, at two values of $Q^2$. 
 
As we go from the NLO to the NNLO analysis, several changes in the  
distributions are worth noting.  First, the decrease of the quark  
distribution at high $x$ and the slight increase at low  
$x$ reflect the behaviour of the coefficient functions $C_{2,q}$
and $C_{2,g}$ respectively.   
Second, recall that in the NLO analysis the input gluon distribution  
decreased at small $x$.  At NNLO we see the gluon  
decreases even more.  This decrease at low $x$ occurs because of the  
increase of $P_{qg}$, see (\ref{eq:a3}).  The consequent rise  
at $x \sim 0.1$ is to ensure that the momentum sum  
rule is satisfied.  The gluon distribution drives the evolution at 
small $x$.  As we evolve to higher $Q^2$, the effect of the NNLO term 
in the splitting function decreases, and a smaller  
gluon leads to slower evolution than in the NLO analysis.  Hence, for  
example, by $Q^2 \sim 10^4$~GeV and $x \sim 10^{-4}$, all NNLO partons 
are about 10--15\% smaller than those at NLO. 
 
Since the biggest NNLO effect is in the small $x$ behaviour of the 
gluon, we study this distribution in more detail.  Fig.~4 shows the  
gluon obtained in the LO, NLO and NNLO global fits at various values 
of $Q^2$.  A clear LO $\rightarrow$ NLO $\rightarrow$  
NNLO hierarchy\footnote{The wobble seen in the LO gluon at $x \sim 
0.1$ for $Q^2 \lapproxeq 20$~GeV$^2$ is a consequence of momentum  
conservation and a much too large a gluon at small $x$.}  of the small 
$x$ behaviour of the gluon is evident, which reflects the  
direct link with the HERA deep inelastic data via $P_{qg}$ of 
(\ref{eq:a3}). Note also that the evolution of the NNLO gluon is made 
even slower because of the (small) negative NNLO  
contribution in $P_{gg}$, see (\ref{eq:a4}). 
 
The \lq starting\rq\ parametric forms of the gluon found in the LO,  
NLO and NNLO global analyses are 
\bea 
\label{eq:a6} 
& & xg (x, Q_0^2) \; = \nonumber \\ 
& & = \; \left \{ \begin{array}{l}  
31.2x^{0.390} (1 - x)^{6.18} \: (1 - 5.23\sqrt{x} + 7.33x) 
\quad\quad\quad\quad\quad\quad\quad\quad\quad\quad\quad\quad\quad\: 
({\rm LO}) \\ 
\label{eq:a7} 
51.8x^{0.535} (1 - x)^{6.55} \: (1 - 3.92\sqrt{x} + 4.68 x) - 1.67x^{-0.032}(1 - x)^{8.21}  
\quad\quad \: ({\rm NLO}) \\ 
\label{eq:a8} 
14.4x^{0.397} (1 - x)^{5.56} \: (1 -3.22\sqrt{x} + 4.36x) - 0.705x^{-0.151}  
(1 - x)^{8.69}.~~({\rm NNLO}) \end{array} \right . 
\eea 
The \lq extreme\rq\ curves $A$ and $B$, plotted in Fig.~4, demonstrate that  
the greatest  
uncertainty, coming from the lack of complete knowledge of the NNLO  
splitting functions, is  
in the small $x$ behaviour of the gluon.  Nevertheless even allowing for  
the \lq extreme\rq\  
spread in the NNLO fits we see that the hierarchy in the small $x$  
behaviour of the gluon persists. 
 
Fig.~4 also shows that the gluon obtained from the NNLO analysis becomes  
negative at small  
$x$ and small $Q^2$, as anticipated in (\ref{eq:a1}).  
However the gluon distribution itself is not a  
physically observable quantity.  It is scheme dependent.  For example, Fig.~4  
shows the  
gluons obtained from analyses at different orders in the  
$\overline{\rm MS}$ factorization  
scheme.  If on the other hand we were to adopt the DIS scheme, then we find  
that the NNLO  
gluon is only marginally negative at low $x$ at $Q^2 = 2$~GeV$^2$.   
In order to investigate  
the true implications of the convergence of the perturbative series we  
must examine the  
predictions for physically observable quantities.  The behaviour of the  
longitudinal structure  
function, $F_L$, is particularly appropriate as it is sensitive to the  
small $x$ behaviour of the  
gluon.  The production cross sections of $W$ and $Z$ 
bosons at the hadron colliders are representative of  
other relevant observables.  We therefore study the predictions for these  
quantities below. 
 
\section{Predictions for $F_L$} 
 
The LO contribution to $F_L$ is ${\cal O} (\alpha_S)$, and so a  
consistent (factorization  
scheme independent) NNLO prediction of $F_L$ requires the  
${\cal O} (\alpha_S^3)$  
coefficient functions.  These are not known at present, but we do know  
much of the same information  
as for the ${\cal O} (\alpha_S^3)$ splitting functions, that is the  
$N = 2, 4, 6, 8$ moments and  
the $x \to 0$ behaviour.  Hence we estimate the coefficient functions in the  
same spirit as used by van Neerven and Vogt for the  
${\cal O} (\alpha_S^3)$ splitting  
functions.  The \lq central\rq\ estimates for the NNLO contributions to  
$C_L$ are (where the common factor of $(\alpha_S/(4\pi))^3$ is taken out) 
\bea 
\label{eq:a9} 
C_{L,g}^{(3)} (x) & = &\Biggl[n_f \biggl( 381 \frac{\ln(1/x)}{x} 
-\frac{1200}{x} +1095\ln^2(1/x) -5960 + 21512x^2  
+ 1928\ln(1-x)\biggl) \nonumber \\ 
&& + n_f^2 \biggl( \frac{20}{x} + 148.8\ln^2(1/x) - 5 -741x^2  
- 147\ln(1-x)\biggr)\Biggr] \\  
\label{eq:a10} 
C_{L,{\rm NS}}^{(3)} (x) & = &\Biggl[\biggl(-323\ln^2(1/x) - 3916 - 47526x^2  
- 21954\ln(1-x)\biggr) \nonumber \\ 
&& + n_f\biggl( -89\ln^2(1/x) + 863 + 2796x^2 + 2038\ln(1-x)\biggr) 
\nonumber \\ 
&& + n_f^2\biggl(15\ln^2(1/x) - 54.3 +72.4x^2 -23\ln(1-x)\biggl)\Biggr]\\ 
\label{eq:a11} 
C_{L,{\rm PS}}^{(3)} (x) & = &\Biggl[n_f \biggl(\frac{169\ln(1/x)}{x} 
-\frac{700}{x} 
+186\ln^2(1/x) + 578x^2 +42.6\ln(1-x) +316\biggr) \nonumber \\ 
&& + n_f^2\biggl(\frac{10}{x} +61\ln^2(1/x) -25+42.7x^2 + 7.2\ln(1-x) 
\biggr) \Biggr] 
\eea 
where NS and PS refer to quark non-singlet and pure-singlet respectively. 
In fact the $n_f^2$ dependent part of the non-singlet coefficient function  
is in principle known exactly from the calculations in \cite{renormalon}, 
but are small and well modelled by our simple analytic expression. 
 
The behaviour of the $F_L$ gluon coefficient function is shown  
in Fig.~5.  The two dominant  
features are (i) a sizeable contribution just below $x = 1$,  
and (ii) a large growth with decreasing $x$ arising from the most singular  
terms found in Ref.~\cite{SX}.  In fact at small $x$ we have\footnote{As for 
$P_{qg}$, at leading order we present the coefficient of the moment space 
coefficient function as $N \to 0$.}  
\be 
\label{eq:a12} 
C_{L,g} (x) \; \simeq \; \frac{\alpha_S}{2\pi} n_f \frac{2}{3} \left [ 
1 - \frac{1}{3}\frac{\bar \alpha_S}{x}  
+ \left (\frac{43}{9} - \zeta(2)\right ) 
\frac{\bar \alpha_S^2 \ln(1/x)}{x} \right ] 
\ee 
and the same expression, modulo the colour factor $C_F/C_A =4/9$,  
for $C_{L,{\rm PS}}$ (except at leading order).   
 
The non-singlet coefficient functions beyond LO are very strongly 
peaked as $x \to 1$. 
At NLO the coefficient function \cite{coefflong} (with $(\alpha_S/(4\pi))^2$  
factored out) in this limit behaves like  
\bea 
\label{eq:a13} 
C^{(2)}_{L,{\rm NS}} (x) & \simeq &  4C_F\biggl[2C_F\ln^2(1-x)+
(9-8\zeta(2))C_F
\ln(1-x) \nonumber \\ 
&& +\biggl(4(\zeta(2)-1)C_A-\bigl(11 -\frac{2}{3}n_f\bigr)\biggr) 
\ln(1-x)\Biggr], 
\eea 
and there is an enhancement compared to the LO result,  
$(\alpha_S/(4\pi))4C_Fx$, due to the $\ln(1-x)$ 
terms. The machinery for computing the dominant $\ln(1-x)$ terms for 
$C_{L,{\rm NS}}$ for all orders in $\alpha_S$ has recently been devised
\cite{STERMAN}, and in principle we could use this to evaluate the parts 
$\propto \ln^m(1-x)$ at ${\cal O}(\alpha_S^3)$ for $m=2,3,4$. However, the
resulting expressions are very far from compact and at this order we simply
choose to use the the information on the moments which is available to give 
us a good estimate of the coefficient function at high $x$. This confirms   
that again the coefficient function is very peaked for $x \to 1$ -- 
its size largely compensating for the extra power of $\alpha_S/(4\pi)$. A more  
sophisticated parameterization than that used in (\ref{eq:a11}) should really 
include higher powers in $\ln(1-x)$, but since the expression matches a range 
of moments very well it will give an accurate representation of the coefficient 
function convoluted with the smooth parton density.   
 
The predictions for $F_L$ obtained from the parton distributions of the  
different global fits  
are shown in Fig.~6.  The progressive increase at high $x$ is attributable  
to the large NS coefficient functions for $x \to 1$.   
At small $x$ the LO and NLO\footnote{Note the very small coefficient of  
$\alpha_S^2/x$ in $C_{L,g}$ of (\ref{eq:a12}).} predictions mirror the  
gluon distribution  
(sampled in the region of $2x$ due to the convolution).  The NNLO prediction  
of $F_L$ also  
mirrors the shape of the gluon at low $Q^2$ and moderate $x$, turning  
over at $x \sim 0.05$.   
Then, at even smaller $x$, the very large ${\cal O} (\alpha_S^3)$  
contribution of $C_{L,g}$  
takes over, which after convolution with the gluon, prevents $F_L$  
becoming negative and,  
in fact, results in a steep rise with decreasing $x$.  As we evolve up  
in $Q^2$ the effect of the  
${\cal O} (\alpha_S^3)$ term in $C_{L,g}$ diminishes and eventually  
the NNLO prediction  
for $F_L$ mirrors the shape and size of the gluon via  
the ${\cal O}(\alpha_S)$ term in  
$C_{L,g}$.  Hence there is a transition at $Q^2 \sim 5~{\rm GeV}^2$ where  
the NLO  
overtakes the NNLO prediction\footnote{In fact at very low $Q^2$  
and $x \sim 10^{-4}$ the  
rate of evolution, $dF_L/d\ln Q^2$, is negative at NNLO.} of $F_L$.   
At the lowest  
values of $Q^2$ the NNLO prediction of $F_L$ should be regarded with  
caution.  If we go  
below $Q^2 = 2$~GeV$^2$ the dip in $F_L$ in Fig.~6 becomes negative,  
indicating the  
unreliability of the NNLO analysis in this domain. 
 
In the region $Q \gapproxeq 20$~GeV$^2$, Fig.~6 shows a LO $\rightarrow$ NLO  
$\rightarrow$ NNLO hierarchy in the small $x$ behaviour of $F_L$,  
which reflects that  
observed for the gluon in Fig.~4.  As compared to the gluon, we see  
that the NNLO effects in  
the $F_L$ coefficient function have improved the stability of the  
predictions somewhat.  The  
degree of stability is displayed in Fig.~7, which shows the NLO/LO and  
NNLO/NLO ratios of  
the $F_L$ predictions for two values of $Q^2$. The convergence is slower  
for small $x$, which most likely is due to the influence of  
missing $\log (1/x)$ terms  
at higher orders.  The convergence improves rather slowly with  
increasing $Q^2$. 
 
\section{Predictions for $W$ and $Z$ hadroproduction} 
 
The cross section predictions for $W$ and $Z$ production at the 
Tevatron  
and the LHC are  
shown in Fig.~8, together with data from the CDF \cite{CDF} 
 and D0 \cite{D0} collaborations. The predictions 
labelled LO, NLO and NNLO are defined (schematically) as follows\footnote{All  
quantities 
are evaluated in the $\overline{\rm MS}$ factorization and  
renormalization schemes, 
with scale choice $Q=M_V$.} 
\begin{eqnarray} 
\sigma_{\scriptstyle\rm  LO} & = & f_{\scriptstyle\rm  LO}  
\otimes f_{\scriptstyle\rm  LO}  \nonumber \\ 
\sigma_{\scriptstyle\rm  NLO} & = & f_{\scriptstyle\rm  NLO}  
\otimes f_{\scriptstyle\rm  NLO}  
\otimes \left[\; 1\; + \; \alpha_{S,{\scriptstyle\rm  NLO}}\; K^{(1)}\; \right] 
\nonumber \\ 
\sigma_{\scriptstyle\rm NNLO} & = & f_{\scriptstyle\rm  NNLO}  
\otimes f_{\scriptstyle\rm  NNLO}  
\otimes \left[\; 1\; + \; \alpha_{S,{\scriptstyle\rm  NNLO}}\; K^{(1)}\;  
+ \; \left( \alpha_{S,{\scriptstyle\rm  NNLO}}\right)^2\; K^{(2)}\;  
\right] 
\label{eq:wz12}
\end{eqnarray} 
where the label on $\alpha_S$ indicates the order to which the  
$\beta-$function is evaluated. The NLO and NNLO contributions $K^{(1,2)}$
are taken from \cite{WILLI}. 
The range of NNLO predictions, corresponding to the $A$ or $B$ choice 
for the approximate NNLO splitting functions, is indicated by the width of  
the band. 
As for $F_L$, the extrema are given by the $AA$ and $BB$ predictions  
(see Eq.~(2)) with the `average' 
NNLO partons giving cross sections very close to the centre of the band. 
Also shown in Fig.~8 (as dashed lines) is the `quasi-NLO' prediction 
\begin{eqnarray} 
\sigma_{\scriptstyle\rm NLO'} & = & f_{\scriptstyle\rm  NLO}  
\otimes f_{\scriptstyle\rm  NLO}  
\otimes \left[\; 1\; + \; \alpha_{S,{\scriptstyle\rm  NLO}}\; K^{(1)}\;  
+ \; \left( \alpha_{S,{\scriptstyle\rm  NLO}}\right)^2\; K^{(2)}\;  
\right] 
\label{eq:wz13}
\end{eqnarray} 
which is the expression used in previous MRST estimates of the $W$ and $Z$  
cross sections \cite{MRST1,MRST2}. The NLO$'$ predictions enable us to  
identify the separate  
NNLO contributions to the cross sections from changing from  
NLO to NNLO partons and  
from including the explicit  ${\cal O}(\alpha_S^2)$ NNLO  
coefficient functions ($K^{(2)}$) in the  
$W,Z$ cross section perturbation series. 
 
The LO  
$\rightarrow$ NLO $\rightarrow$ NNLO convergence of the predictions is  
much better than  
for $F_L$, because the boson cross sections depend mainly on the  
{\it quark} distributions at $x  
\sim 0.05$ (Tevatron) and $x \sim 0.006$ (LHC).  Since the global fits  
include high precision  
$F_2$ data, there is considerable stability in  
the quark distributions in the sampled $x$  
regions, see Fig.~3.   
 
The jump from $\sigma_{\scriptstyle\rm  LO}$ to  
$\sigma_{\scriptstyle\rm  NLO}$ is 
 mainly due to the well-known large ${\cal O}(\alpha_S)$ double logarithmic  
Drell-Yan K--factor correction arising from soft-gluon emission. 
The NLO and NNLO cross sections are much closer. By comparing with  
the NLO$'$ predictions,  
we see that at the Tevatron energy the increase of 
about $+4\%$ from NLO to NNLO is due in roughly equal parts to the slight  
increase in the  
$u$ and $d$ partons in this $x$ range (see Fig.~3), and the net  
effect of the various $K^{(2)}$ 
contributions. 
 
At the LHC energy the NLO and NNLO predictions are even closer, 
because  
(a) the  
$K^{(2)}$ contribution is smaller due to an almost complete  
cancellation between the positive 
$q \bar q$ and negative $qg$ contributions \cite{WILLI}, and (b)  
the quark ratios average to unity at $x \sim 0.006$ for $Q^2 \sim  
10^4~{\rm GeV}^2$, see Fig.~3.  The NNLO band is larger than at the Tevatron  
because the partons are probed at smaller $x$, where there is more uncertainty  
in the NNLO evolution.  
 
We may conclude from Fig.~8 that perturbative convergence is not a  
dominant uncertainty in predicting the $W$ and $Z$ cross sections.   
This stability indicates the potential value of these processes acting as a  
luminosity monitor for the Tevatron and the LHC. 
 
\section{Conclusions} 
 
In this paper we have taken a first look at a NNLO global parton analysis of  
deep inelastic and  
related hard scattering data.  Although the NNLO splitting functions are  
not fully known,  
enough information is available to bound their possible behaviour.   
Even allowing for the full  
spread of the uncertainties of the functions, we are able to draw  
interesting conclusions.  The  
inclusion of NNLO effects gives an overall improvement in the description  
of the data, which  
is due to the increased scaling violations at both large and small $x$.
In a similar manner, if higher-twist contributions are allowed, they 
decrease in magnitude for both large and small $x$ as we increase the 
order, approaching 
very small values for $x \lapproxeq 0.3$, but remaining large and positive 
at large $x$. The latter behaviour largely reflects the expectations
arising from the presence of heavy target corrections. 

Fitting to the data using  
LO, NLO and NNLO frameworks leads to a hierarchy of gluon distributions at  
small $x$,  
such that the NNLO ($\overline{\rm MS}$) input gluon is found to go  
negative for $x \lapproxeq 10^{-3}$. 
However, we stressed that perturbative convergence should be tested for  
physical observables, rather  
than for the parton distributions themselves. 
To this end, the LO, NLO and NNLO  
predictions were made  
for the longitudinal structure function $F_L$, and for $W$ and  
$Z$ hadroproduction cross-sections.   
Although the input gluon goes negative for $x \lapproxeq 10^{-3}$,  
we found that $F_L$ is  
positive for $Q^2 \gapproxeq 1~{\rm GeV}^2$.  Despite this the form of
the predictions for $F_L$ show that the DGLAP approach is  
not convergent  
until $Q^2 \sim 5$~GeV$^2$.  The convergence then improves slowly with  
increasing $Q^2$ and  
reveals a LO $\rightarrow$ NLO $\rightarrow$ NNLO hierarchy  
in the predictions for $F_L$,  
which mirrors that of the gluon but with increased stability.  A measure  
of the uncertainty is  
the $\sim 15\%$ change in $F_L$ in going from the NLO to NNLO  
prediction at $x \sim  
10^{-3}$ and $Q^2 \sim 100~{\rm GeV}^2$.  The convergence  
deteriorates with decreasing  
$x$ and most likely is due to the neglect of $\log (1/x)$  
contributions beyond the NNLO  
DGLAP framework.  At low $Q^2$ ($Q^2 \lapproxeq 5~{\rm GeV}^2$)  
the $\log (1/x)$  
terms are even more important. There is also the possibility  
of higher-twist contributions, which for $F_L$ may be different at
small $x$ from those for $F_2$ \cite{BARTELS}.   
On the other hand the predictions of the $W$ and $Z$ hadroproduction  
cross sections are  
rather stable, due to the more direct relation between the fitted data  
and the predictions. 
 
Here we have addressed, in an exploratory fashion, theoretical issues  
arising from including  
NNLO corrections in global parton analyses of deep inelastic  
and related data.  However new 
HERA data with increased precision will soon be available.   
These will be included in a new  
global analysis to yield both an updated set of NLO partons and a  
first set of NNLO distributions. 
 
\section*{Acknowledgements} 
 
We thank Willy van Neerven and Andreas Vogt for providing us  
with compact analytic expressions for the NNLO splitting functions, and 
Andreas Vogt also for useful discussions. This work was supported in part by
the EU Fourth Framework Programme ``Training and Mobility of Researchers'',
Network ``Quantum Chromodynamics and the Deep Structure of Elementary
Particles'', contract FMRX-CT98-0194 (DG 12 - MIHT).   
 
\section*{Appendix :  NNLO treatment of heavy flavour partons} 
 
For the treatment of heavy flavours we use an approximate NNLO generalization 
of the Thorne-Roberts variable flavour number scheme  (VFNS).  
This scheme was 
presented in detail in \cite{TR}, and the general framework outlined for 
all orders in perturbation theory. Essentially one obtains the VFNS  
coefficient functions in terms of the fixed flavour number scheme 
(FFNS) coefficient functions and partonic matrix elements $A_{ab}$.
The former are the coefficient functions calculated assuming that the heavy
quark (denoted by $H$) has no parton distribution, but may only be created 
via a hard scattering process. The matrix elements 
define the $(n_f+1)$--flavour parton distributions in terms of the 
$n_f$--flavour parton distributions, i.e. the $A_{Ha}$ tell one how the heavy
quark distribution is constructed from the light partons and the $A_{ab,H}$ 
tell one how the light parton distributions 
are altered by internal heavy quarks
(in particular $A_{ab,H}^{(0)} = \delta_{ab}$). 
The VFNS coefficient functions are
determined by solving Eqs. (3.5)-(3.9) in the latter of 
\cite{TR}. For example, 
\be 
\label{eq:ap1} 
C^{{\rm FF}(n)}_{Hg} = \sum_{m=0}^n C^{{\rm VF}(n-m)}_{Hg} \otimes 
A^{(m)}_{gg,H} 
+ n_f C^{{\rm VF,PS}(n-m)}_{Hq} 
\otimes A^{(m)}_{qg,H} + [C^{{\rm VF,NS}(n-m)}_{HH} +C^{{\rm VF,PS}(n-m)}_{HH}]
\otimes A^{(m)}_{Hg}. 
\ee  
The matrix 
elements and FFNS coefficient functions are unambiguously calculable, 
but there 
is some element of choice in the VFNS coefficient functions since there are 
more degrees of freedom than there are constraining equations. One may 
eliminate this ambiguity by simply calculating diagrams assuming one 
has initial 
state heavy partons and keeping mass dependent terms. However, this leads to  
unphysical threshold behaviour for the coefficient functions, and we choose 
instead to impose as physical a constraint as possible. Hence,  we make the 
derivative of $F_2^H(x,Q^2)$ continuous in the gluon sector (which 
overwhelmingly dominates) as one switches from FFNS to VFNS coefficient 
functions and turns on the heavy quark parton distribution at 
$Q^2=m_H^2$. This choice of VFNS coefficient functions is essentially 
a freedom in factorization schemes, with all schemes becoming identical when 
summed to all orders, but differing by terms $\sim m_H^2/Q^2$ at 
finite order. 
 
At NNLO, all VFNSs experience two related technical complications due 
to internal quark loops which may or may not 
be cut. First, it has long 
been known that the parton distributions become discontinuous at $\mu^2=m_H^2$  
at ${\cal O}(\alpha_S^2)$ \cite{Buza}.\footnote{This discontinuity 
begins at ${\cal O}(\alpha_S)$ in some factorization schemes.} For 
example, the heavy quark distribution at $\mu^2 =m_H^2$ becomes 
\be 
\label{eq:ap2} 
(H+\bar H)^{(2)} \: (x,m_H^2) = \biggl(\frac{\alpha_S(m_H^2)}{2\pi}\biggr)^2 
[A^{(2)}_{Hg}\otimes g(m_H^2) + A^{(2)}_{Hq}\otimes q (m_H^2)]. 
\ee 
The gluon and light quarks also acquire discontinuities as the heavy 
parton distribution is turned on, such that momentum is conserved, see \cite{Buza}. These  
lead to a corresponding 
discontinuity in the coefficient functions, maintaining the continuity of the  
structure functions, e.g. solving (\ref{eq:ap1}) at NNLO at 
$\mu^2=m_H^2$ one obtains 
\be 
\label{eq:ap3} 
C^{{\rm FF}(2)}_{Hg} = C^{{\rm VF}(2)}_{Hg} + C^{{\rm VF,NS}(0)}_{HH}  
\otimes A^{(2)}_{Hg}. 
\ee 
The second complication at NNLO arises because the heavy quarks in the final 
states are no 
longer just those coupling directly to the external vector boson probe,  
but can be generated even when it is a light quark coupling to this 
probe. In principle it is a  
technical shortcoming of our scheme that the implicit definition of the heavy 
quark structure function involves the heavy quark coupling  
to the external vector boson. This simplifies the factorization, but is not 
strictly physically correct. A more general prescription is discussed in  
\cite{CNS}, where a cut in invariant mass has to be implemented above  
which heavy quark-antiquark 
pairs generated away from the external vertex may be defined as observable.  
 
In this paper we simply ignore both these complications. This is due 
to the fact 
that the whole analysis is approximate and also because both  
lead to effects which in practice are extremely small\footnote{The 
change in parton 
distributions across threshold was investigated in \cite{CNS}, but using GRV98 
NLO parton distributions \cite{GRV}. The discontinuity is dominated by 
the gluon at small 
$x$. The GRV gluon at small $x$ is large at $\mu^2=m_c^2$, while ours 
is small and even becomes 
negative at the same scale. Hence, the effect is very much smaller.}  
-- especially when compared 
to other uncertainties. Both complications should be dealt with in a truly 
precise NNLO analysis once the exact NNLO splitting functions are 
known, though we are confident that they (especially the latter) will 
lead to tiny effects. However, at present we do not even know the 
NNLO, i.e. ${\cal O}(\alpha_S^3)$, 
FFNS coefficient functions, so a precise VFNS is impossible to define.  
 
Nevertheless, this is where our heavy flavour prescription comes into  
its own. Other prescriptions \cite{ACOT,CNS} which use the coefficient 
functions from diagrams 
involving single initial state heavy partons rely on precise cancellations  
between the heavy quark distributions and terms involving the VFNS coefficient  
functions in order to maintain smooth behaviour. For example, at 
NNLO a large contribution to the heavy quark evolution from  
$\alpha_S^3 P^{(2)}_{qg}$ needs to be cancelled by a term  
$\alpha_S^3 P^{(2)}_{qg} \ln(\mu^2/m_H^2)C^{{\rm VF}(0)}_{2,HH}
\otimes g(\mu^2)$  
to avoid too quick a growth of $F_2^H(x,Q^2)$ for $\mu^2$ just above $m_c^2$. 
In our prescription the correct threshold behaviour is built into  
$C^{{\rm VF}(0)}_{2,HH}$ 
automatically, i.e. $C^{{\rm VF}(0)}_{2,HH}=0$ if $W^2 < 4 m_H^2$, and such  
precise cancellations are not necessary 
-- simply including NNLO evolution of the heavy quarks 
without NNLO heavy quark coefficient functions at all maintains smooth 
behaviour. However, we want to obtain the correct NNLO high $Q^2$ 
limit. Hence, we  
include the massless ${\cal O}(\alpha_S^2)$ coefficient functions for 
the heavy quarks, but  
weighted by a factor of $\beta = (1-4m_H^2z/(Q^2(1-z))^{0.5}$,  
i.e. the velocity of the 
heavy quark in the centre of mass system, to impose the correct 
threshold behaviour at low $Q^2$.  
This procedure is very simplistic, but it contains all the relevant physics.  
Significant improvements to this approximate 
procedure would require the NNLO FFNS coefficient functions. 
 
Finally, we note that the heavy flavour longitudinal coefficient 
functions behave like $\beta^3$, and thus are heavily suppressed until very  
high $Q^2$. At such high $Q^2$, the ${\cal O}(\alpha_S^3)$ 
coefficient functions  
have become relatively unimportant, and hence we simply omit the  
${\cal O}(\alpha_S^3)$ longitudinal coefficient functions until a more 
precise analysis is possible.

\newpage 

\begin{figure}[H]
\begin{center}
\epsfig{figure=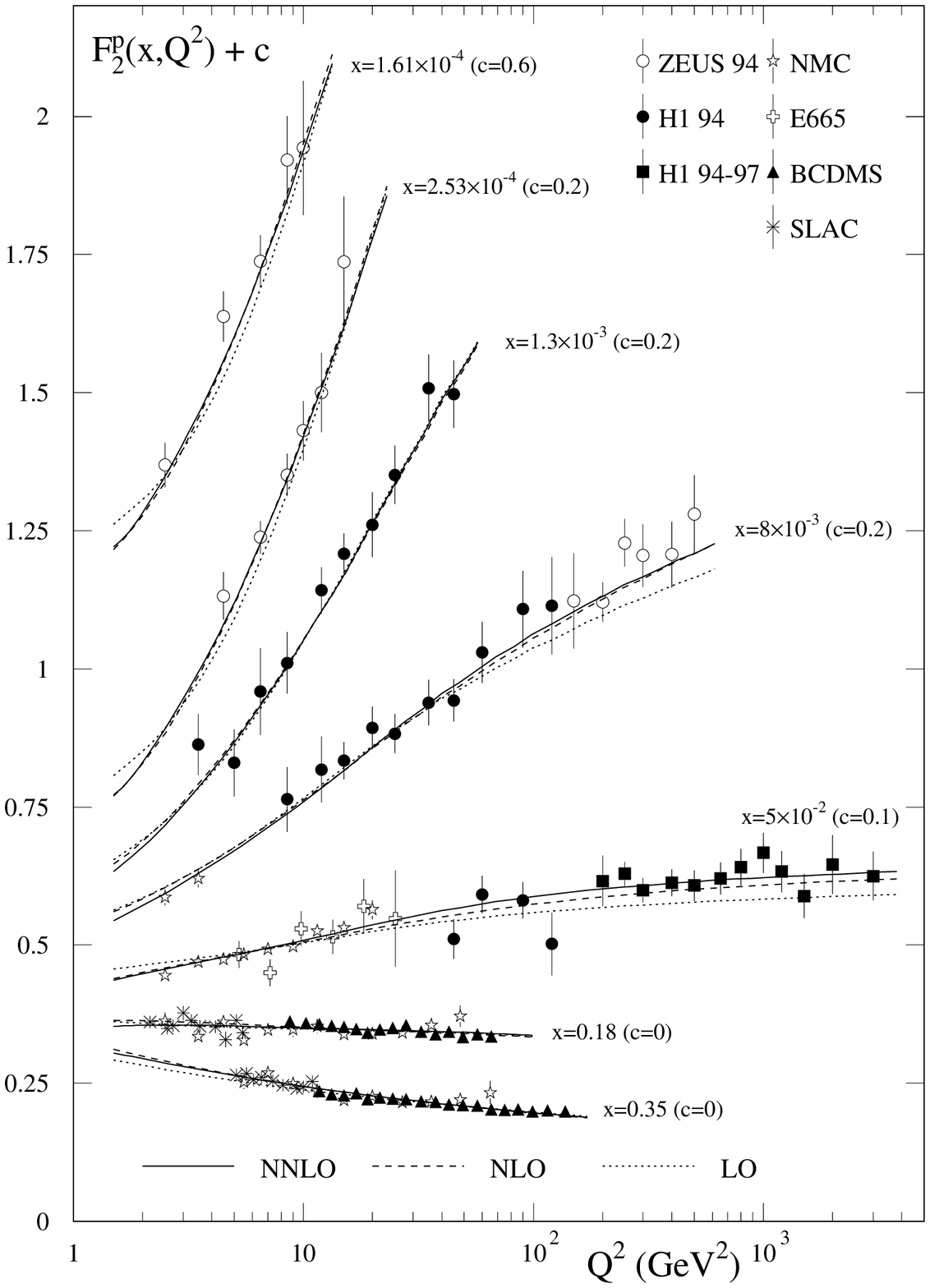,height=20cm}
\end{center}
\caption{The description of data \protect\cite{F2} for the $F_2$ 
structure function at a few representative $x$ values obtained in the 
LO, NLO and NNLO global parton analyses.} 
\label{fig:Fig1}
\end{figure}

\begin{figure}[H]
\begin{center}
\epsfig{figure=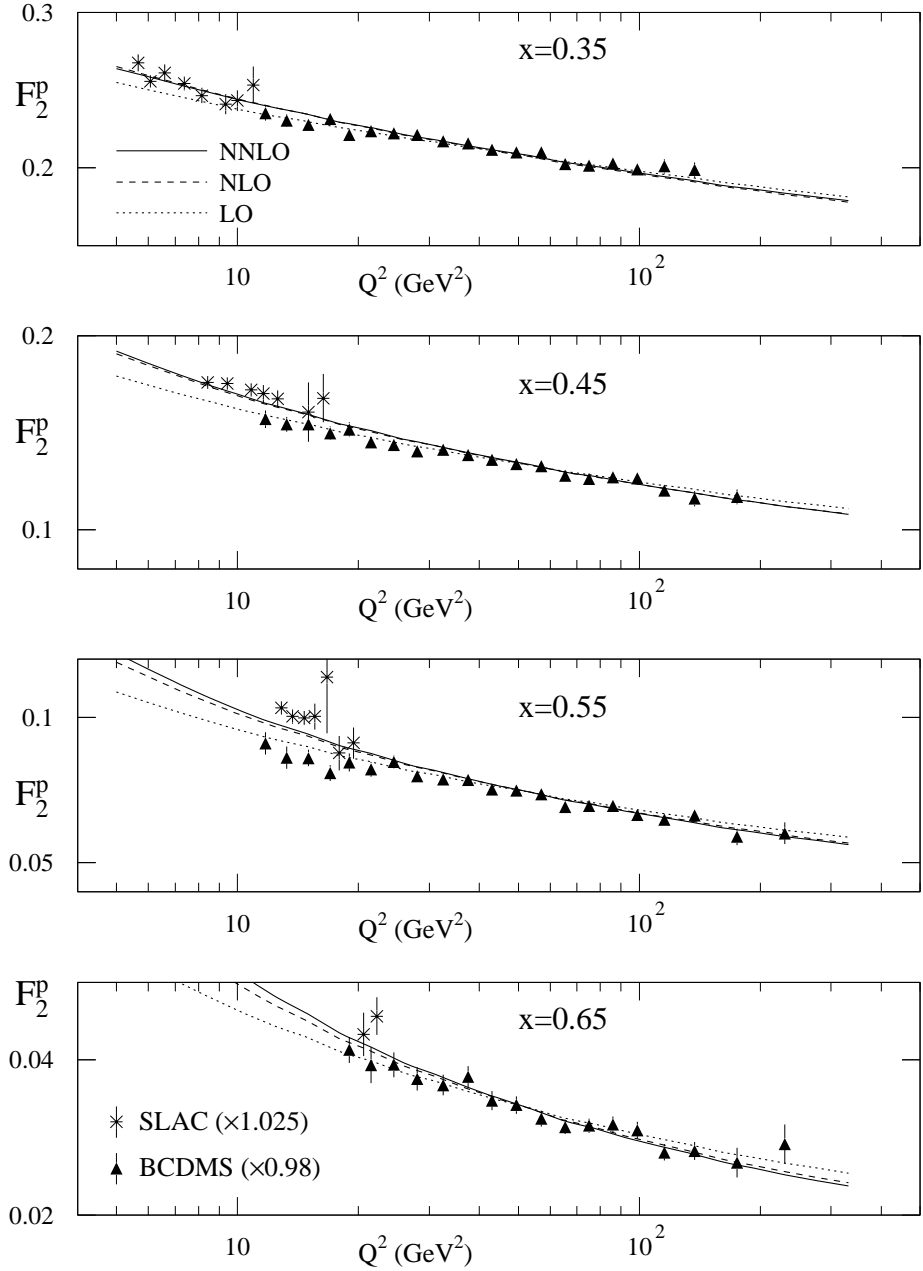,height=20cm}
\end{center}
\caption{The description of data \protect\cite{F2} for the $F_2$ 
structure function at large $x$ obtained in the 
LO, NLO and NNLO global parton analyses.} 
\label{fig:Fig2}
\end{figure}

\begin{figure}[H]
\begin{center}
\epsfig{figure=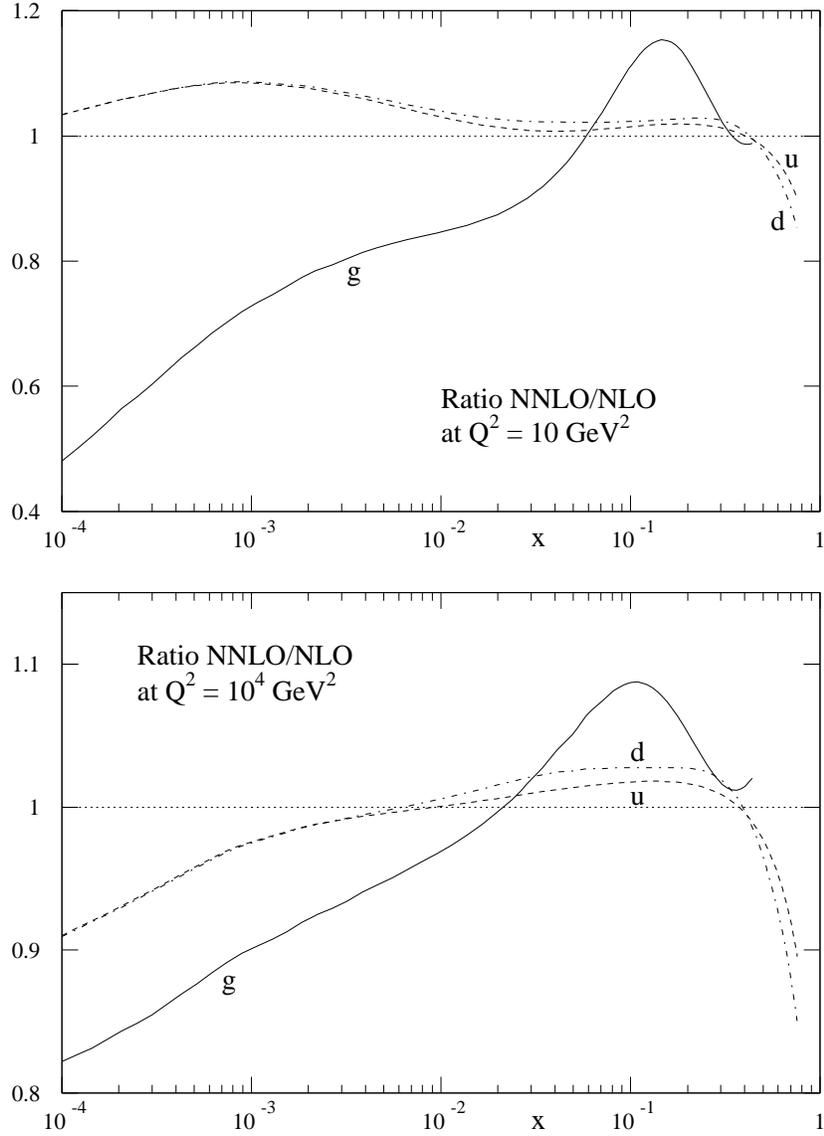,height=18cm}
\end{center}
\caption{A comparison of partons obtained in the \lq central\rq\  
NNLO analysis with those obtained in the NLO fit, first at $Q^2 =  
10$~GeV$^2$ and then at $Q^2 = 10^4$~GeV$^2$.  We show the NNLO/NLO  
ratios for the gluon and the up and down quark distributions.} 
\label{fig:Fig3}
\end{figure}

\begin{figure}[H]
\begin{center}
\epsfig{figure=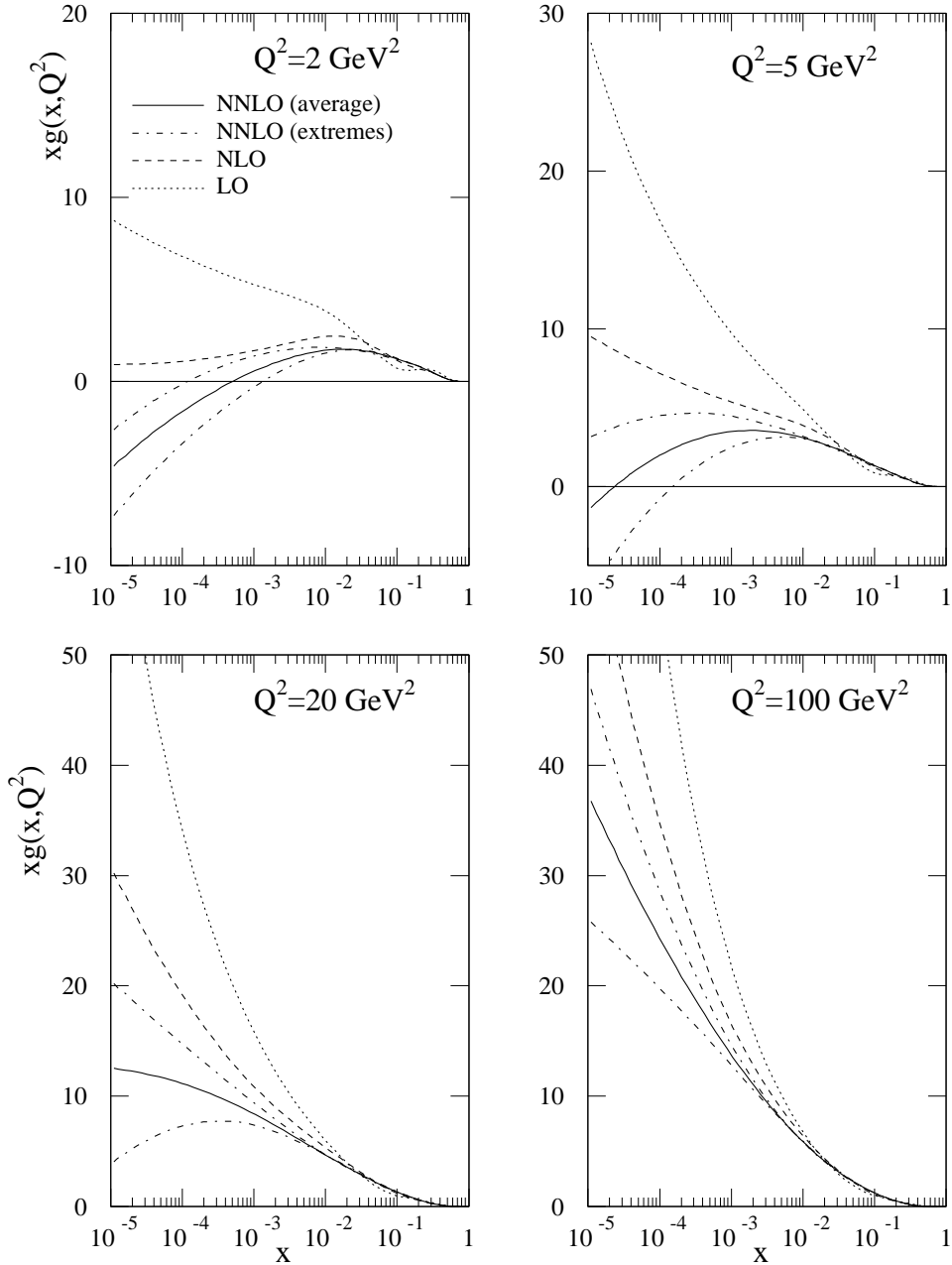,height=20cm}
\end{center}
\caption{The evolution of the gluon obtained in the LO, NLO and  
NNLO global analyses.  The gluons obtained using the extreme forms, 
$A$ and $B$, of the NNLO splitting  
functions are shown (dot-dashed curves), together with that from the 
average (continuous curves).} 
\label{fig:Fig4}
\end{figure}

\begin{figure}[H]
\begin{center}
\epsfig{figure=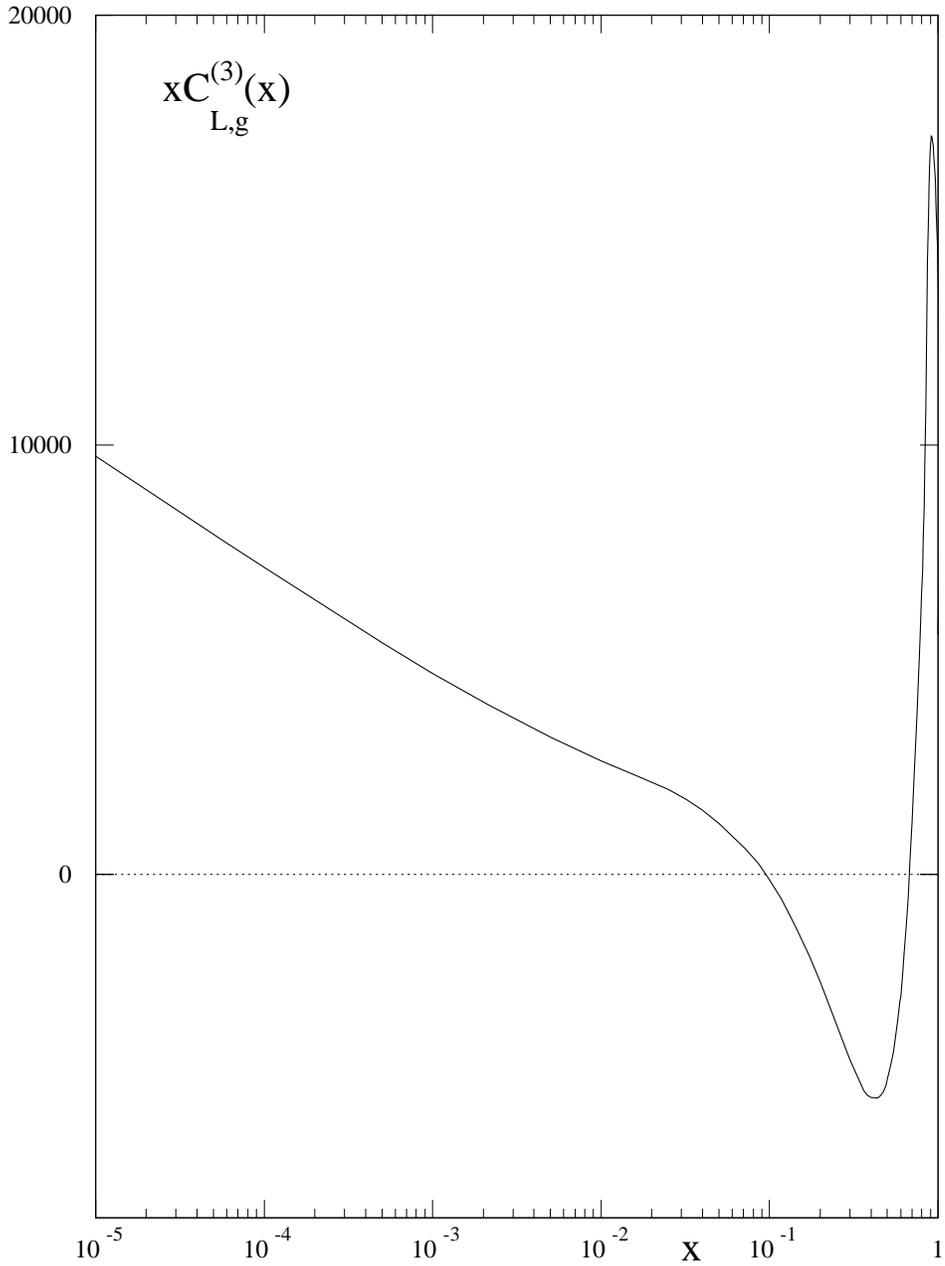,height=20cm}
\end{center}
\caption{The behaviour of the NNLO contributions to the  
coefficient function $xC^{(3)}_{L,g}(x)$ for  
$F_L$ taking $n_f=3$.  The average of the two extreme behaviours is shown.} 
\label{fig:Fig5}
\end{figure}

\begin{figure}[H]
\begin{center}
\epsfig{figure=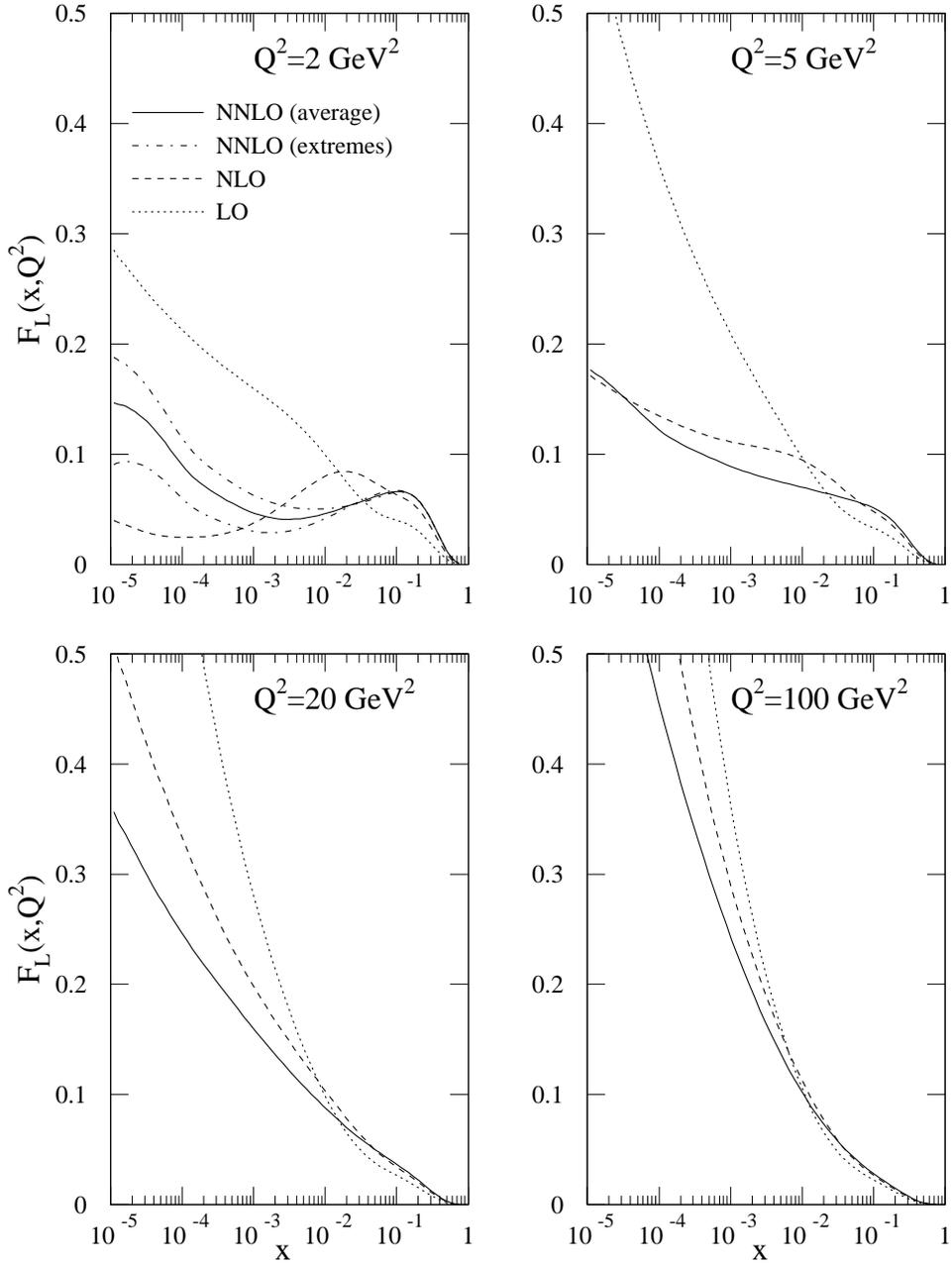,height=20cm}
\end{center}
\caption{The predictions for the longitudinal structure  
function $F_L$ obtained from the LO, NLO and NNLO sets of partons.
The NNLO extremes (using parton uncertainty alone) 
are plotted only for $Q^2 = 2$ GeV$^2$.} 
\label{fig:Fig6}
\end{figure}

\begin{figure}[H]
\begin{center}
\epsfig{figure=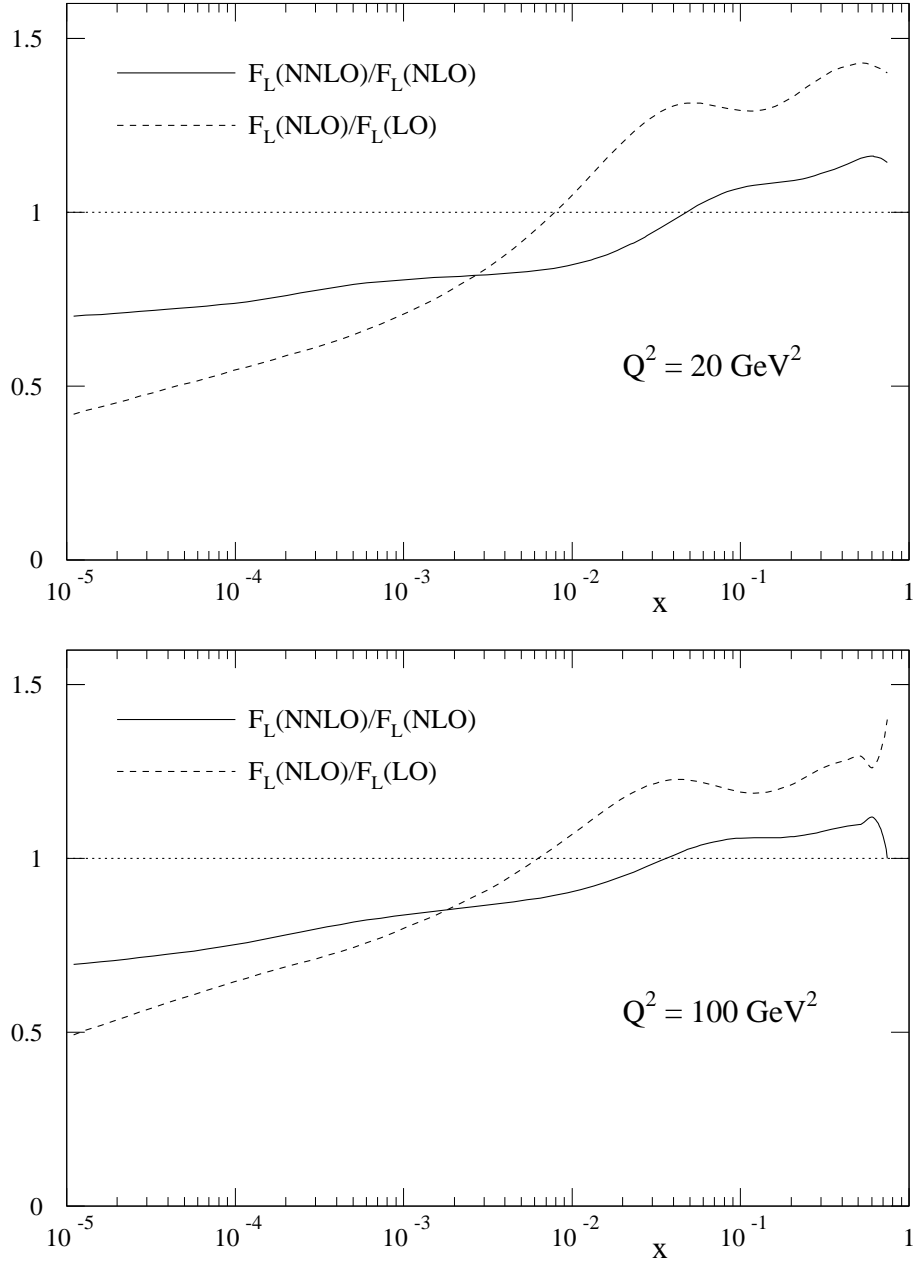,height=20cm}
\end{center}
\caption{The NLO/LO and NNLO/NLO ratios of the predictions of $F_L$,  
at two different values of $Q^2$,  
shown to indicate the degree of perturbative stability of the analysis.} 
\label{fig:Fig7}
\end{figure}

\begin{figure}[H]
\begin{center}
\epsfig{figure=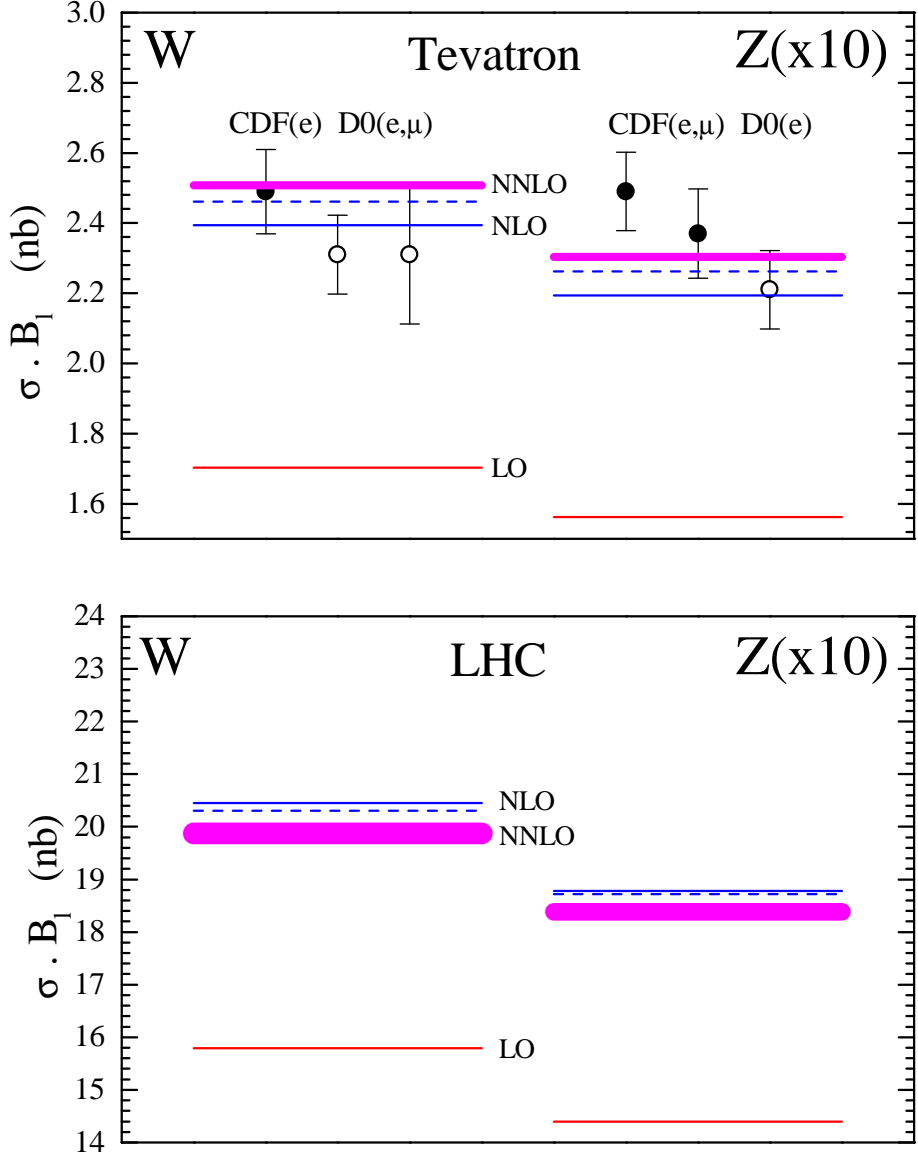,height=20cm}
\end{center}
\vspace{-2.0cm}
\caption{The predictions of the cross sections for $W$ and $Z$  
production and leptonic decay 
at the Tevatron and the LHC obtained from parton sets of the LO,  
NLO and NNLO global analyses.   
The cross sections labelled LO, NLO, NLO$'$ (dashed line) and NNLO are  
defined in Eqs. (\ref{eq:wz12},\ref{eq:wz13}). The band of 
NNLO predictions corresponds to the $A,B$ variation of the small--$x$  
approximate splitting functions, as discussed in the text.  
Also shown are measurements  
obtained at the Tevatron \protect\cite{CDF,D0}. We take the leptonic branching
ratios $B(W \to l\nu) = 0.1084$ and $B(Z \to l^+l^-) = 0.03364$.} 
\label{fig:Fig8}
\end{figure}

\end{document}